# COMPARATIVE CLINICAL EVALUATION OF "MEMORY-EFFICIENT" SYNTHETIC 3D GENERATIVE ADVERSARIAL NETWORKS (GAN) HEAD-TO-HEAD TO STATE OF ART: RESULTS ON COMPUTED TOMOGRAPHY OF THE CHEST

Mahshid Shiri[1], Chandra Bortolotto[2], Alessandro Bruno[3], Alessio Consonni[4], Daniela Maria Grasso[4], Leonardo Brizzi[4], Daniele Loiacono[1], Lorenzo Preda[2]

[1] Dipartimento di Elettronica, Informazione e Bioingegneria, Politecnico di Milano, Milan, Italy
[2] Radiology Institute, Fondazione IRCCS Policlinico San Matteo Pavia - Università degli studi di Pavia, Italy
[3] Department of Business Law Economics, Consumer Behavior "Carlo A. Ricciardi", Faculty of Communication, IULM University, Milan, Italy
[4] Radiology Institute, Fondazione IRCCS Policlinico San Matteo Pavia, Italy

## ABSTRACT

**Introduction**: Generative Adversarial Networks (GANs) are increasingly used to generate synthetic medical images, addressing the critical shortage of annotated data for training Artificial Intelligence (AI) systems. This study introduces CRF-GAN, a novel memory-efficient GAN architecture that enhances structural consistency in 3D medical image synthesis. Integrating Conditional Random Fields (CRFs) within a two-step generation process, allows CRF-GAN improving spatial coherence while maintaining high-resolution image quality. The model is designed to be computationally efficient, avoiding the need for additional GANs or post-processing. Its performance is evaluated against the state-of-the-art hierarchical (HA)-GAN model.

**Materials and Methods**: We evaluate the performance of CRF-GAN against the state-of-the-art hierarchical (HA)-GAN model. The comparison between the two models was made through a quantitative evaluation, using Fréchet Inception Distance (FID) and Maximum Mean Discrepancy (MMD) metrics, and a qualitative evaluation, through a two-alternative forced choice (2AFC) test completed by a pool of 12 resident radiologists, in order to assess the realism of the generated images.

**Results**: CRF-GAN outperformed HA-GAN with lower FID (0.047 vs. 0.061) and MMD (0.084 vs. 0.086) scores, indicating better image fidelity. The 2AFC test showed a significant preference for images generated by CRF-Gan over those generated by HA-GAN with a p-value of 1.93e-05. Additionally, CRF-GAN demonstrated 9.34% lower memory usage at $256^3$ resolution and achieved up to 14.6% faster training speeds, offering substantial computational savings.

**Discussion**: CRF-GAN model successfully generates high-resolution 3D medical images with non-inferior quality to conventional models, while being more memory-efficient and faster. The key objective was not only to lower the computational cost but also to reallocate the freed-up resources towards the creation of higher-resolution 3D imaging, which is still a critical factor limiting their direct clinical applicability. Moreover, unlike many previous studies, we combined qualitative and quantitative assessments to obtain a more holistic feedback of model's performance.

## 1 Introduction

Artificial Intelligence (AI) systems are proving to be valuable tools for supporting radiologists in the diagnostic process. They may be especially helpful in the interpretation of medical images, improving diagnostic accuracy and the timeliness of therapeutic interventions [2]. However, the success of these systems and their widespread implementation in clinical settings is still limited by the availability of large datasets with annotated medical data, which are essential for their training [3]. Annotated medical data from experts is often difficult to obtain due to privacy concerns, high costs, complexity of collection procedures, and the need for approval by ethical committees [4]. In light of these challenges, the expansion of datasets with synthetic labeled data is of increasing interest to researchers and clinicians.

Previous models for dataset expansion were represented by Data Augmentation systems, such as intensity transformations, rotation, translation, cropping, and other geometric transformations [4,5]. However, these methods had limitations in capturing the heterogeneity of real image distributions, as they generated images through a one-to-one process. In this context of scarcity of quality data, Generative Adversarial Networks (GANs) [6] offer a promising solution. By learning the characteristics and distributions of real data, GANs can generate realistic and entirely new medical images that replicate the variability and complexity of images used in clinical practice. This enables the training of more robust and accurate AI systems. GANs, whose classic architecture was proposed in 2014, consist of two deep neural networks: the Generator and the Discriminator. These two networks operate simultaneously in a minimax game. The Generator creates new data samples, while the Discriminator evaluates both generated and real samples, aiming to distinguish between them. This ongoing competition drives both components to improve their performances.

3D GAN models have been proposed for various applications, including reducing noise in low-dose CT scans [7], enhancing the quality of CT images [8], generating realistic 3D brain MRI images [9], and creating tumor masks for segmentation [10]. In the segmentation field, Dong et al. [11] proposed an adversarial training strategy to train deep neural networks for the segmentation of multiple organs on thoracic CT images, with the aim of improving the efficiency of chest radiotherapy treatment planning. Shubhangi et al. [12] proposed RescueNet, an unpaired adversarial training to segment brain tumor. In a recent study, Najafi et al. [13] achieved excellent results in lung segmentation and subsequent tumor detection in the segmented areas by combining the GAN, Long Short-Term Memory, and VGG16 algorithms. Chen et al. [14] obtained good results in segmentation of lung lesions and blood vessels, achieving realistic three-dimensional reconstruction and visualization results. Swaminathan et al. [15] address the critical challenge of diagnosing lung cancer at later stages by leveraging deep learning to enhance early detection, especially for individuals at high risk. GANs have been widely proposed also in data augmentation, with the aim of improve limited amount of data, which is also one of the goal of our work. Frid-Adar et al. [16] used a Deep Convolutional GAN (DCGAN) to generate liver lesions of three labeled types (cysts, metastases, hemangiomas) in order to augment the initial dataset size of a liver lesion classifier. With the aim of a more comprehensive generation of the lung structure to allow the development of models to capture more information about lung cancer, Mendes et al. [4] proposed generated CT chest images from corresponding positional and semantic annotations using two generative adversarial networks and databases of real computed tomography scans: the Pix2Pix and the Conditional Generative Adversarial Network. In addition, Onishi et al. [17] used generated CT images for pulmonary nodules classification, developing a pretraining method which used GAN for improving the DCNN classification performance of pulmonary nodules. Lot of studies on GANs have employed cross-modality translations, for example, generating CT-like images based on MR images or generating MR images across different sequences (for instance, T1-weighted to T2-weighted). Jin et al [18] used GAN to transform brain CT images to MR images also for radiotherapy planning. Hong et al. [19] proposed a GAN-LSTM-3D method for 3D reconstruction of lung cancer tumors from 2D CT images, proving that GANs can be a good option to segment lung and tumor.

In the specific area of CT image generation, Ferreira et al. [20] and Han et al. [5] ,aimed to synthesize realistic 3D images of lung parenchyma or lung nodules using different 3D GAN models: Progressive Growing (PG)-GAN in [20] and a Multi-Conditional (MC)-GAN in [5]. However, these methods have a limitation in the generated image resolution, which is usually $128^3$ or smaller, because of limited memory during training [21]. Memory-efficient GANs seek to balance memory efficiency and generative capability, ensuring effective GAN training even with limited computational resources. To achieve this, Lei et al. [22] and Yu et al. [23] focused on generating new images slice-by-slice or patch-by-patch. However, as these methods generate patches and slices independently, potentially leading to artifacts at the boundaries. Also, Uzunova et al. [24] uses two GANs for image generation, with the first network producing a lower-resolution version and the second generating higher-resolution patches conditioned on the first GAN's output. This approach remains patch-based and lacks a comprehensive understanding of the entire image structure. Furthermore, Sun et al. [21] introduced an end-to-end hierarchical GAN architecture (HA-GAN) capable of generating high-resolution 3D images at a resolution of $256^3$. HA-GAN comprises two interconnected GANs: a low-resolution GAN that produces a low-resolution version of the 3D image, capturing the essential global structure with reduced computational and memory demands, and a high-resolution GAN that generates high-resolution patches for a randomly selected sub-volume of the image. HA-GAN [21] has shown superior performance compared to several baseline models, including WGAN [30], VAE-GAN [31], αGAN [32], ProgressiveGAN [33], 3D StyleGAN 2 [34], and CCE-GAN [35]. Additionally, HA-GAN is the only model that can generate images at a resolution of $256^3$, overcoming memory constraints that limit other models. For this reason, our proposed architecture is compared directly to HA-GAN.

These approaches have shown the potential of GANs to generate high-quality, realistic medical images, often aiding in diagnostic processes and improving imaging techniques. However, the majority of these works have relied on deep and complex GAN architectures that demand considerable computational resources. These models typically involve large numbers of parameters and extensive training data, which results in a high demand for processing power and memory. While these methods have proven successful in terms of output quality, their computational cost can limit their practical application, especially in environments with constrained hardware or where real-time processing is essential.

Our goal was to propose a GAN model with reduced computational capacity in order to address the high computational demands of previous methods, while still achieving high-quality results. By simplifying the architecture of the GAN, we aimed to reduce the need for extensive processing power, making the model more efficient and feasible for real-world applications. The key objective was not only to lower the computational cost but also to reallocate the freed-up resources towards the creation of higher-resolution 3D imaging. This way, we could focus more on the quality of the 3D rendering, enabling the generation of detailed and precise models without sacrificing the performance of the system as a whole. The proposed architecture combines Conditional Random Fields (CRFs) with GANs to reduce memory usage while improving performance. CRF is a probabilistic graphical model that models dependencies between output variables, considering the sequential or structural nature of the data. CRFs enable the capture of correlations between image patches. The unary potential in a CRF represents the likelihood of assigning a particular label to a patch based solely on its features, while the pairwise potential considers the relationship between neighboring patches, encouraging label consistency in adjacent regions and capturing the spatial structure and dependencies within the image.

In the field of medical imaging, especially in generative model research for tasks like CT image synthesis, evaluating the generated images is crucial for determining the model's effectiveness. Many past studies have predominantly used quantitative metrics [2][5] such as structural similarity, pixel-wise accuracy, or various loss functions to assess the quality of generated images. While these metrics provide objective, data-driven insights into how closely the model's output matches the real data, they often overlook aspects like visual realism or clinical authenticity that are vital in medical applications. In our study, we introduced an important qualitative evaluation component, carried out by radiology residents with expertise in chest CT imaging. This is significant because human experts are able to assess not just the numerical resemblance of the generated images to real scans, but also their clinical accuracy, such as the correct representation of anatomical structures, disease patterns, and overall image fidelity. These are aspects that quantitative metrics alone cannot fully capture. This combination of quantitative and qualitative evaluation represents a key step forward in medical imaging research, providing a more balanced and clinically relevant assessment of generative models.

The primary contribution of this work is to assess the clinical realism of synthetic images generated by memory-efficient GANs. This process is crucial for validating the practicality and applicability of these images in real-world medical settings, thereby enhancing the potential of AI-driven healthcare solutions. The structure of this work may be summarized as follows:

1. We present a novel, memory-efficient architecture by incorporating CRFs into the middle layers of GANs to capture anatomical structures, aiming to increase consistency in the synthetic images. We evaluate the computational performance of our model in terms of model complexity and memory efficiency.

2. We assessed the clinical value of the synthetic CRF-GAN images by comparing them, through qualitative and quantitative evaluation, with synthetic images from another GAN model (HA-GAN) used as a "competitor."

3. We evaluate benefits and limitations of our model and compare it to other GAN models.

## 2 Materials and Methods

### 2.1 Datasets

The study utilizes a publicly accessible 3D dataset employed for the Lung Nodule Analysis 2016 (LUNA16) challenge [27] which is a subset of the LIDC-IDRI dataset [28]. The Lung Image Database Consortium image collection (LIDC-IDRI) constitutes diagnostic thoracic computed tomography (CT) scans annotated with marked lesions. It serves as an internationally accessible resource for the development, training, and evaluation of AI systems for lung cancer detection and diagnosis. Initiated by the National Cancer Institute and further advanced by the Foundation for the National Institutes of Health, this is public-private partnership, accompanied by the Food and Drug Administration. This dataset comprises 1,018 diagnostic thoracic computed tomography (CT) scans, each annotated by four experienced radiologists who identified and categorized lesions into nodules ≥3 mm, nodules <3 mm, and non-nodules [28].

A subset of the LIDC/IDRI database, publicly accessible, was utilized for the LUNA16 challenge [27]. CT scans with a slice thickness exceeding 2.5 mm were excluded, resulting in a total of 888 cases for analysis. The entire dataset was partitioned into 10 subsets, accessible as compressed zip files. Within each subset, CT images are stored in MetaImage (mhd/raw) format.

The dataset was divided into two subsets, allocating 90% (800 cases) for training and 10% (88 cases) for validation. The LUNA16 dataset was chosen due to the following advantages: (i) It provides high-quality annotations verified by four radiologists, ensuring reliability and paving the way for future directions of generating images conditioning on the annotations; (ii) It is publicly accessible, facilitating reproducibility and benchmarking in research; and (iii) All participants involved in the qualitative assessment phase were experts in Chest CT imaging, enhancing the evaluation reliability.

During the preprocessing phase, in line with the approach of [21], blank axial slices are eliminated by substituting them with zero values. Subsequently, the images are resized to dimensions of $256^3$. Additionally, the Hounsfield units (HU) of the images undergo calibration, and air density correction is applied. To ensure uniformity, the HU are mapped to the intensity window of [-1024, 600], and normalized to the range [-1, 1]. The images were resized to dimensions of $256^3$ voxels to standardize the input size for the model.

## 2.2 Model

Model Description:

The proposed architecture introduces a novel two-step GAN framework, using CRFs for improved image synthesis by emphasizing consistency and structural coherence. GANs typically refine an image progressively across their layers, with initial layers generating coarse structures and later layers fine-tuning details. The generator in this architecture is divided into two sequential components: the first component (G1) generates an embedding that encapsulates the image's global structure, with the second (G2) refining the embedding into high-resolution image patches. The second component processes only a subset of the embedding during training to achieve memory efficiency, generating patches of the image rather than the entire output. This subset is selected by a random index (r) drawn from a uniform distribution. This selective approach reduces computational overhead. During inference, the second component of the generator utilizes the complete embedding to produce a full-resolution image, ensuring that the final output is both structurally consistent and visually detailed. A key innovation in this approach is the integration of CRFs to analyze and enhance correlations between different regions of the image embedding. Unlike prior methods (21) that require additional GAN models for structural consistency, CRFs provide a *lightweight mechanism* to improve inter-patch coherence without significantly increasing computational cost.

The architecture also incorporates a "half-encoder" component (hE), which extracts embeddings from image patches and compares them to their generated counterparts during training. This component reinforce the generator's consistency and prevents mode collapse. This encoder takes image as input and generates embeddings that are compared to the original ones during training. By focusing on embedding correlations, the CRF enhances the generator's ability to produce structurally coherent and realistic images. During training, a dual-feedback mechanism operates: the CRF ensures structural alignment within the embeddings, while the discriminator evaluates the realism of the output image patches. This lightweight and memory-efficient approach improves the generator's ability to synthesize consistent and high-quality images. So the Overall Loss function is formulated as Equation 1, where $L_{GAN}$ is the typical GAN loss, $L_{CRF}$ is the Binary Cross Entropy Loss that is calculated based on CRF feedback, and $L_{reconstruction}$ aligns embeddings via the half-encoder by calculating $l_1$ distance.

$$L_{Total} = L_{GAN} + L_{CRF} + L_{reconstruction}$$

$$L_{GAN} = \min_{G_1,G_2} \max_{D} \left[ \mathbb{E}_{r \sim p_r} \left[ \mathbb{E}_{x \sim p_{data}} [\log D(x_{r:r+constant})] \right] \right] + \mathbb{E}_{z \sim p_z} [\log(1 - D(G_2(G_1(z)_{r:r+constant})))]$$

$$L_{CRF} = \min_{G_1,G_2} \max_{CRF} \left[ \mathbb{E}_{r \sim p_r} \left[ \mathbb{E}_{x \sim p_{data}} [\log CRF(hE(x)_{r:r+constant})] \right] \right] + \mathbb{E}_{z \sim p_z} [\log(1 - D(G_1(z)))]$$

$$L_{reconstruction} = \min_{hE} [\mathbb{E}_{x \sim p_{data}, r \sim p_r} ||X_{r:r+constant} - G_2(hE(X_{r:r+constant}))||_1] \quad (1)$$

The two-step GAN structure and CRF integration present a significant advance in generative modeling, offering a lower computational footprint compared to traditional methods. The overall architecture is illustrated in Figure 1.

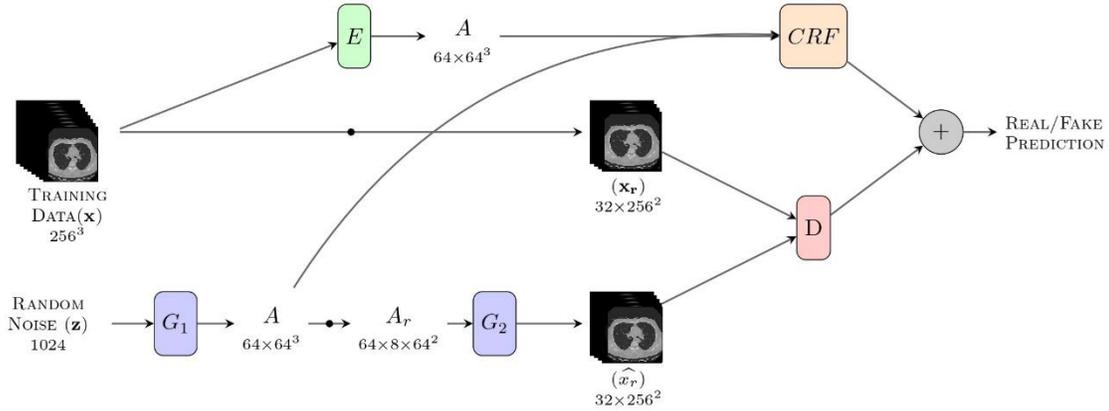

Figure 1 CRF-GAN Architecture.

**Discussion:** Our approach distinguishes itself from previous methods by leveraging a CRF component. In earlier studies [24] [21], researchers addressed the challenge of producing consistent 3D images during inference by introducing an auxiliary GAN tasked with generating low-resolution images. This extra task was designed to enforce general consistency across image patches in the feature domain but at the *cost of increased complexity* and *reduced memory efficiency*. In contrast, our method directly incorporates a *lightweight* CRF component. CRFs are a type of probabilistic graphical model that excel in capturing dependencies between neighboring image patches. Instead of requiring an auxiliary task, the CRF evaluates each patch individually while also considering the relationships between adjacent patches. It does this by assigning a score to each patch based on its features and by comparing neighboring patches to ensure that similar patches receive consistent labels. This approach naturally enforces spatial coherence and maintains the overall structure of the image *without the added computational burden*.

The proposed CRF-GAN architecture was implemented using PyTorch, incorporating 3D CNN layers for image synthesis. The model was trained for 80 epochs on a TITAN V GPU with 12 GB of memory. The Adam optimizer was employed for training due to its adaptive learning rate properties. The learning rate for generator, discriminator, CRF and half-encoder are set as 1e-4, 4e-4, 1e-4 and 1e-4 respectively. The training process utilized binary cross-entropy loss for binary classification tasks. Additionally, L1Loss was employed for the encoder, promoting sharper and more detailed outputs compated to L2. The discriminator's architecture integrated spectral normalization to stabilize GAN training, while the generator leveraged transposed convolutional layers with batch normalization and ReLU activation. The encoder employed 3D convolutional layers and group normalization. The hyperparameters were the same as [21].

The flowchart of preprocessing, training and inference phases of our work is presented in Figure 2.

### 2.3 Evaluation

The comparison with HA-GAN has been conducted from both a quantitative and qualitative perspective, while the performance evaluation also considers key parameters such as maximum memory usage, the number of learnable parameters, and training speed. The evaluation procedures employed are detailed below.

#### 2.3.1 Quantitative Evaluation

To evaluate the synthetic images generated by two models, we used two key metrics: Fréchet Inception Distance (FID) and Maximum Mean Discrepancy (MMD).

Heusel et al. [37] introduced the Fréchet Inception Distance (FID) as a metric for assessing the quality of generated samples. FID accomplishes this by embedding a set of generated samples into a feature space provided by a specific layer of Inception Net or any Convolutional Neural Network (CNN). The embedding layer is treated as a continuous multivariate Gaussian, and both the mean and covariance are estimated for the generated and real data. The Fréchet distance, also known as the Wasserstein-2 distance, between these two Gaussian distributions is then utilized to quantify the quality of the generated samples. This metric is calculated as in Equation 2, where $\left(\mu_{r_{data}}, \Sigma_{r_{data}}\right)$ and $\left(\mu_{g_{data}}, \Sigma_{g_{data}}\right)$ are mean and covariance of real and generated data respectively.

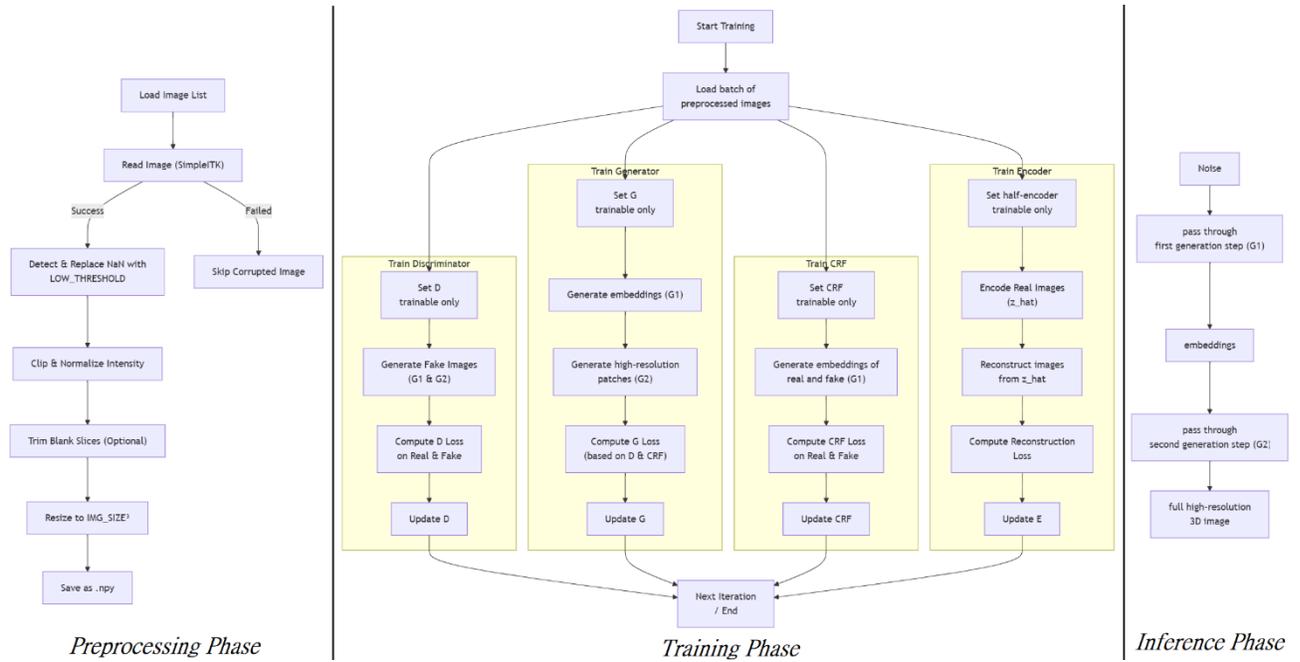

Figure 2 The flowchart of Preprocessing, Training and Inferences phases.

$$FID(r_{data}, g_{data}) = \left\Vert \mu_{r_{data}} - \mu_{g_{data}} \right\Vert_2^2 + Tr(\Sigma_{r_{data}} + \Sigma_{g_{data}} - 2\sqrt{\Sigma_{r_{data}}\Sigma_{g_{data}}}) \quad (2)$$

A lower FID value indicates smaller distances between the synthetic and real data distributions. FID is recognized for its ability to discriminate between samples, its robustness, and its computational efficiency. Although FID is considered a reliable measure, it assumes that the features follow a Gaussian distribution, which may not always hold true. The FID metric degrades as various types of artifacts are introduced into the images. This score aligns well with human judgments and focuses on measuring the dissimilarity between the generated and real distributions.

The measure known as MMD calculates the difference between two probability distributions by using independently drawn samples from each distribution [38]. A lower MMD value indicates that two distributions are closer. MMD can be considered as a two-sample testing method (i.e. Distinguishing two distributions by finite samples), which tests whether one model or another is closer to the true data distribution. MMD score is calculated as Equation 3 where P and Q are the two probability distributions being compared, x, y are the samples drawn from each distribution and k(x,y) is the kernel function (e.g. Gaussian) used.

$$MMD(X, Y) = E_{x,x' \sim P}[k(x, x')] + E_{y,y' \sim Q}[k(y, y')] - 2E_{x \sim P, y \sim Q}[k(x, y)] \quad (3)$$

Overall, FID measures how closely the generated images match the real image distribution, accounting for both precision and recall. MMD, on the other hand, assesses the difference between real and generated images with minimal sample and computational complexity. Lower FID and MMD values indicate that the generated images closely resemble real images.

### 2.3.2 Qualitative Evaluation

The qualitative evaluation of images aims to assess the clinical value of the synthetic CRF-GAN images by evaluating their realism from a purely subjective standpoint rather than using objective metrics. The evaluation was performed by a pool of twelve (12) resident radiologists, working at IRCCS San Matteo of Pavia, with experience dealing with chest CT imaging. The majority of them (9/12, 75%), at the time of evaluation, were attending the fourth and last year of residency, one of them (8,34%) was attending the third year of residency and two of them (16,67%) were attending the second year of residency. All the residents involved already completed the chest CT internal rotation of six months, assuming that they were equally experienced on that specific topic and could provide evaluation on thoracic imaging with the same amount of expertise. Every resident reviewed all the proposed cases. They use the same monitor, that was calibrated to DICOM GSDF. The evaluation was conducted through a two alternative forced choice (2AFC) test composed of two sections, for a total of 40 pairs of images:

1. In the first section, the residents evaluated 10 pairs of images; each couple was composed by a real image and a generated image, for a total of 20 images. The evaluator was presented with a total of 10 questions each presenting two sets of randomly selected images from a chest CT scan, acquired at the same slice (position) on the same anatomical plane (axial, coronal or sagittal). One of the two images had been artificially generated using CRF-GAN model, while the other was a real CT, resized at a $256^3$ resolution to match the resolution of the generated image, but no information about their origin was provided.

Table 1: FID and MMD scores computed on the $256^3$ resolution images of Genomics Superstruct Project and LUNA16 dataset. As lower FID and MMD scores outline better image fidelity, it can be seen that CRF-GAN achieved better results.

| Models | LUNA16 dataset | |
|---|---|---|
| | FID ↓ | MMD ↓ |
| HA-GAN | 0.061021 | 0.086461 |
| CRF-GAN | **0.047062** | **0.084015** |

   The task of the evaluator was to identify which image was the real one. Subsequently, the evaluator had to assess the level of difficulty of the task using a Likert scale from 1 (extremely subtle) to 5 (obvious).

2. In the second section, the residents evaluated 30 pairs of images; each couple was composed by an image generated by HA-GAN and another one generated by CRF-GAN, for a total of 60 images. The evaluator was presented with 30 questions displaying two randomly selected slices of chest CT images each, acquired at the same position and on the same anatomical plane; both images of each pair had been artificially generated, one from CRF-GAN model and the other one from HA-GAN model. The evaluator's task was to indicate the most realistic one.

The test had to be completed in one sitting as it was not possible to pause and resume later to avoid any bias.

To determine the statistical significance of the preference for CRF-GAN or HA-GAN through our study, we used chi-square test.

### 2.3.3 Computational Performance Evaluation

For computational performance evaluation, we used a test set consisting of 10% of the LUNA16 dataset, with all experiments conducted at a resolution of $256^3$. LUNA16 (Lung Nodule Analysis 2016) is a 3D medical imaging dataset consisting of CT scans from the LIDC-IDRI dataset, with a specific focus on lung nodule detection. To evaluate the model's complexity and efficiency, we monitored the maximum memory usage during training with batch sizes of 2, 4, and 6. Furthermore, to enhance our evaluation, we quantified the total number of learnable parameters within the models to assess their complexity, as a higher number of parameters necessitates a larger dataset. This consideration is particularly significant in clinical applications since medical datasets are often limited in size compared to those of natural images. Additionally, we measured the training speed of the models by conducting training over 1000 iterations and determining the average number of iterations completed per second.

## 3 Results

### 3.1 Quantitative results: FID and MMD scores

Lower FID and MMD scores serve as indicators of the proximity of generated images to real images, reflecting the performance of the models. These scores are calculated on $256^3$ resolution of the synthetic images produced by CRF-GAN and HA-GAN and are presented in Table 1, with CRF-GAN outperforming HA-GAN in terms of both metrics. For this evaluation we used 2048 synthetic images of each model.

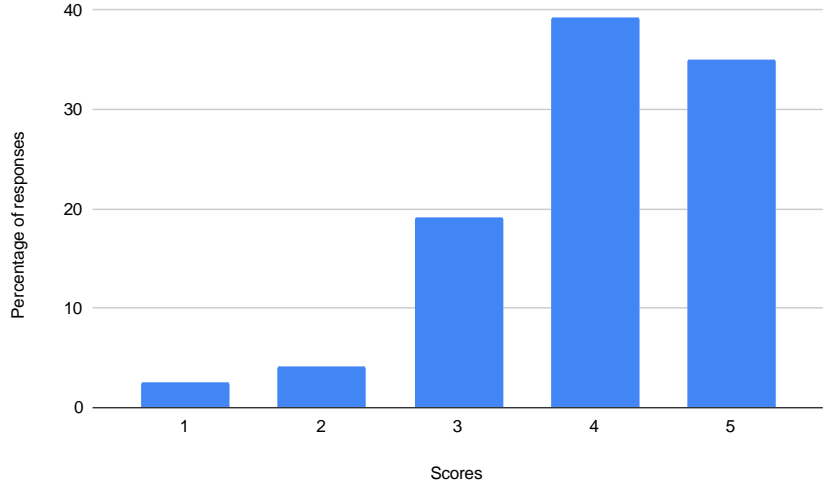

Figure 3: The graph displays the difficulty scores for identifying real and synthetic images. The y-axis shows the percentage of responses for each score, while the x-axis represents the range of difficulty scores with 1: 'Extremely Subtle', 2: 'Moderately Subtle', 3: 'Fairly Subtle', 4: 'Moderately Obvious' and 5: 'Obvious'.

### 3.2 Qualitative results: 2AFC test

In the first section of the test, we compared real and synthetic images. All twelve experts correctly identified all real images for every question which shows that the synthetic images are not identical with real ones. Analyzing the difficulty ratings reveals that synthetic images rarely posed challenges for the experts as can be seen in Figure 3.

This study also aimed to compare the performance of two synthetic image generation models, CRF-GAN and HA-GAN. Figure 4 presents examples of image pairs, illustrating that in some instances HA-GAN was preferred more frequently, while in others, CRF-GAN received a higher number of selections. The study focused on quantifying participant preferences and analyzing the statistical significance of these preferences between the two models. Descriptive statistics in Table 2 provide an overview of the total votes received by each model, as well as the average and variability in votes per image pair.

Table 2: Descriptive Statistics of Votes

| Statistic | Value |
| --- | --- |
| Total Votes for CRF-GAN | 215 |
| Total Votes for HA-GAN | 145 |
| Mean Votes for CRF-GAN per Pair | 7.16 |
| Mean Votes for HA-GAN per Pair | 4.83 |
| Standard Deviation of Votes for CRF-GAN | 2.66 |
| Standard Deviation of Votes for HA-GAN | 2.66 |
| Chi-square (p-value) | 71.42 (1.93e-05) |

### 3.3 Computational Performance Evaluation

#### 3.3.1 Model Complexity:

We explored the complexity of CRF-GAN and HA-GAN models by analyzing the total number of learnable parameters at different resolutions, which are presented in Table 3. The proposed model demonstrates a reduction in parameters at each resolution, with decreases of 27.4%, 28.1%, and 28.2% observed for resolutions $64^3$, $128^3$, and $256^3$ respectively, as highlighted in the table.

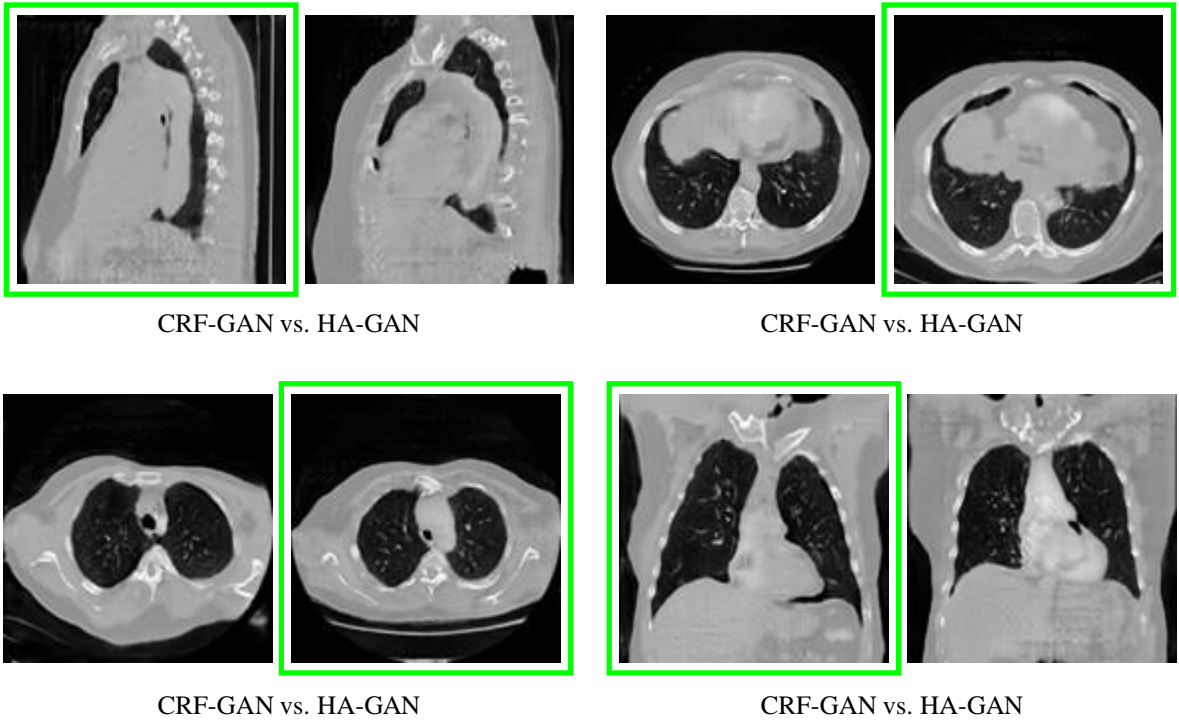

| | | | |
|---|---|---|---|
| CRF-GAN vs. HA-GAN | | CRF-GAN vs. HA-GAN | |
| CRF-GAN vs. HA-GAN | | CRF-GAN vs. HA-GAN | |

Figure 4: Sample pairs from the 2AFC test. The image pair that was chosen more often as realistic is bounded with a green box (CRF-GAN on the left, HA-GAN on the right).

In addition, we measured the number of iterations per second during training, setting the batch size to 2. The comparison was conducted at $128^3$ and $256^3$ resolutions. The results indicate that at a $256^3$ resolution, CRF-GAN achieves a training speed of 1.394 iterations per second, whereas HA-GAN achieves 1.216 iterations per second. This means CRF-GAN is approximately 14.6% faster than HA-GAN at this resolution. At a $128^3$ resolution, CRF-GAN achieves 4.236 iterations per second, while HA-GAN achieves 3.845 iterations per second, making CRF-GAN about 10.2% faster at this lower resolution.

### 3.3.2 Memory efficiency:

In order to compare memory usage, we captured the maximum memory allocated by each model during training. This procedure was performed on batch sizes of 2, 4, and 6 on both $128^3$ and $256^3$ resolutions. The results are outlined in Table 4.

As it can be seen, HA-GAN consistently demanded more memory resources than CRF-GAN across different resolutions and batch sizes. On average, HA-GAN utilized approximately 9.65% more memory compared to CRF-GAN at $128^3$ resolution and approximately 9.34% more memory at $256^3$ resolution.

Table 3: The parameter counts for CRF-GAN and HA-GAN at varying resolutions, which are indicated in millions (M) in the table.

| Resolution | #Parameters | |
|---|---|---|
| | CRF-GAN | HA-GAN |
| $64^3$ | 57.07 M | 78.7 M |
| $128^3$ | 57.20 M | 79.58 M |
| $256^3$ | 57.24 M | 79.74 M |

Table 4: Memory usage (in MB) of HA-GAN and CRF-GAN for different batch sizes at $128^3$ and $256^3$ resolutions. "–" indicates that the models exceed GPU memory.

| Models | $128^3$ | | | $256^3$ | | |
|---|---|---|---|---|---|---|
| | 2 | 4 | 6 | 2 | 4 | 6 |
| **HA-GAN** | 2812 | 2984 | 3098 | 7710 | 10434 | – |
| **CRF-GAN** | 2622 | 2712 | 2774 | 7494 | 9010 | – |

### 3.4. Hyperparameter Selection and Scalability:

The hyperparameters in our approach are derived from heuristics used in state-of-the-art methods, without any fine-tuning specific to our dataset. We conducted additional experiments on a larger brain imaging dataset with a significantly higher number of samples to demonstrate that these hyperparameters are not dataset-specific and to evaluate the scalability of our method. For this purpose, we used GSP dataset containing meticulously curated cross-sectional data derived from 1,570 individuals aged between 18 and 35. The results show that our method maintains strong performance, with CRF-GAN achieving an FID of 0.007044 and an MMD of 0.000099, outperforming HA-GAN (FID: 0.009278, MMD: 0.003802). These findings confirm the generalizability and robustness of our approach across different datasets.

## 4 Discussion

Memory-efficient 3D GANs may facilitate data generation while utilizing minimal computational resources, thereby expanding datasets to mitigate data scarcity; these models, increasing iterations and so image quality without increasing generation time, may not only save time and resources but also uphold a high standard of data quality. High-resolution computed tomography (CT) scans play a crucial role in clinical diagnostics, offering radiologists a powerful tool to make informed decisions. This is especially important when examining small anatomical structures or pathological changes, such as in the detection of lung nodules. Lot of previous study in the field of generated images were focused on the characterization of lung cancer [5] [11] [19]. This study aims to enhance the understanding of lung cancer by focusing on the detailed generation of lung structures, enabling the development of models that can capture more nuanced and comprehensive information. By leveraging generative models in this context, researchers can create more accurate and complex representations of lung anatomy, which can, in turn, aid in the detection and analysis of pathological features. These innovative generative approaches can be used to simulate various lung conditions, incorporating concrete clinical findings into the models for improved predictive accuracy.

Previous works [5] [24] rely on intricate and computationally heavy GAN models, which require substantial processing power and memory. These models typically involve a vast number of parameters and demand large datasets for training, resulting in a high computational cost. While they deliver impressive image quality, their resource-intensive nature can limit their practical use, particularly in environments where computational resources are limited or when real-time processing is critical. In our knowledge only Sun et al. [21] introduced a low computational and memory demanding high-resolution 3D GAN architecture outperforming state-of-art models that generate coherent high-resolution 3D images, called HA-GAN; for this reason, our proposed architecture is compared directly to HA-GAN. In this work, we proposed a novel memory-efficient architecture for high-resolution 3D medical image synthesis thanks to the implementation of Conditional Random Fields and we assessed the clinical value of the synthetic images by comparing them, through qualitative and quantitative evaluation.

From a computational performance and workload point of view, CRF-GAN proved to be a lighter model than HA-GAN, both in terms of learnable parameters and memory usage. The significantly fewer learnable parameters of CRF-GAN compared to HA-GAN, leads to higher computational efficiency during training, reflecting in a larger number of iterations per second at both $128^3$ and $256^3$ resolution. Regarding memory usage, HA-GAN consistently demanded more memory resources than CRF-GAN across different resolutions and batch sizes. Interestingly, as the batch size increased, this percentage difference tended to increase at both resolutions, indicating that higher batch sizes amplify the disparity in memory usage between the two models. By simplifying the architecture of the GAN, we aimed to reduce the need for extensive processing power, making the model more efficient and feasible for real-world applications. The key objective was not only to lower the computational cost but also to reallocate the freed-up resources towards the creation of higher-resolution 3D imaging.

In terms of quantitative evaluation, both FID and MMD metrics values show better image fidelity for CRF-GAN.

Quantitative metrics are essential for objectively assessing the quality of images generated by synthesis models, as they measure the discrepancy between the distributions of real and synthetic images. However, relying solely on these metrics can be misleading due to their inherent limitations. For instance, FID assumes a Gaussian distribution of the features extracted from the images, which is not always true, thereby compromising the reliability of the evaluation. Similarly, MMD quantifies the difference between two probability distributions but does not provide information on which specific features contribute to this difference, limiting the interpretability of the results. Therefore, it is crucial to complement these metrics with qualitative assessments to adopt a holistic approach that allows for a deeper understanding of the quality of the generated images.

In medical imaging research, particularly when developing generative models for tasks like CT image synthesis, the evaluation of generated images plays a crucial role in determining the effectiveness of the model. Traditionally, many studies [13] [17] have focused on quantitative metrics to evaluate the quality of generated images. These metrics are valuable because they provide objective, numerical measures of how well the model's output approximates the target or ground truth. However, they often fail to capture more subjective, qualitative aspects of image realism that are critical for medical applications. In our study, we introduced a qualitative evaluation of generated images performed by experienced radiology residents with expertise in chest CT scans. The rationale behind this is that human experts can assess the clinical plausibility of generated images, looking at features like anatomical accuracy, tissue differentiation, and realistic depiction of disease patterns, all aspects that are difficult to measure through traditional quantitative methods. Our findings show that the doctors evaluations deemed the images produced by our model to be more realistic than those from the HA-GAN model, which highlights the clinical relevance and potential utility of our approach in real-world settings. The importance of qualitative evaluation cannot be understated, especially when considering the application of these models in clinical practice. In terms of qualitative assessment process, the 2AFC test allowed radiologists residents to formulate some subjective, but globally shared, evaluations on the clinical nature of the images, stemming from their expertise in the field of medical radiology. The qualitative difference, evident in comparison with real images, is not evident in comparison with the generated ones: The correctness of this assumption is confirmed by the results of the 2AFC test. The result of the 2AFC test shows a clear preference for CRF-GAN over HA-GAN. Out of the total votes, CRF-GAN received 215 (59,7%), while HA-GAN garnered 145, indicating that participants were more likely to choose images from CRF-GAN as appearing more realistic. On average, CRF-GAN received 7.16 votes per image pair, compared to HA-GAN's 4.83. Both models showed similar variability in votes with a standard deviation of 2.66. A chi-square test was performed to assess the statistical significance of this preference. The test yielded a chi-square statistic of 71.42 with an extremely low p-value of 1.93e-05, demonstrating that the observed difference is highly unlikely to be due to random chance. This indicates a strong and statistically significant preference for CRF-GAN over HA-GAN in terms of generating more realistic images.

However, there remains a significant gap between the current capabilities of GAN-generated images (HA- or CRF-) and their practical application in real-world clinical settings as the qualitative distance between generated and real images is still evident, both in terms of spatial resolution and anatomical accuracy. While the qualitative difference is undeniable, it is yet to be determined whether a poor performance from a purely clinical or subjective perspective correlates with limitations in quantitative metrics. This question can guide future studies aimed at evaluating the performance of AI systems trained on medical datasets, while also exploring the potential of using 3D GAN models for data augmentation to improve diagnostic accuracy.

There are some limitations in the current study:

- Participant Bias in Medical Image Evaluation: participant bias could be introduced, as individual criteria for evaluating image quality vary among the participants, potentially affecting the outcomes. In medical imaging, the evaluation of image quality is often influenced by factors such as the evaluator's level of expertise, their interpretation of certain features, or even their personal preferences and experience. These differences can cause disparities in how image quality is rated, potentially affecting the outcomes of studies or evaluations, and may result in findings that are not entirely reflective of the true performance of the system being studied.

- Low number of participants: The number of radiology residents who evaluated the generated images (twelve) may restrict the generalization of the results. Increasing the number of participants in future studies is essential for improving the reliability and generalizability of the results. A larger sample size allows researchers to better account for variability and differences among individuals, ensuring that the findings are not limited or biased by a small, homogenous group.

- The Need for Pathological Findings: we evaluated only the GAN's capability of generating chest CT volumes with no pathological findings. However, since AI systems require training on pathological features, future studies should explore the GAN's ability to generate and accurately represent pathological findings, such as lung nodules. This exploration is crucial for developing models that can assist in the diagnosis of conditions like lung cancer, where accurate detection of pathological features is essential.

- Resolution of generated images: The resolution at which the images were generated ($256^3$) represents a step forward from the point of view of 3D generative models, as already discussed in previous sections. However, in the clinical-radiological field, this resolution does not reflect the one usually used, which is normally $512^3$ or higher, depending on the CT scanner setting used. Furthermore, it should be noted that this aspect did not represent a bias in the comparison of the generated images with real ones since the real images were downscaled to a $256^3$ resolution, making them comparable with the generated ones. There remains a significant gap between the current capabilities of GAN-generated images (HA- or CRF-) and their practical application in real-world clinical settings as the qualitative distance between generated and real images is still evident, both in terms of spatial resolution and anatomical accuracy. While the qualitative difference is undeniable, it is yet to be determined whether a poor performance from a purely clinical or subjective perspective correlates with limitations in quantitative metrics. This question can guide future studies aimed at evaluating the performance of AI systems trained on medical datasets, while also exploring the potential of using 3D GAN models for data augmentation to improve diagnostic accuracy.

- 2D Slice Evaluation in 3D GAN-Generated Chest CT: although both CRF-GAN and HA-GAN are 3D models, the quality evaluation by residents has been conducted on 2D slices randomly selected from the generated volume. This procedure doesn't follow the usual CT scan interpretation method in an everyday clinical context, made of scrollable volumes. However, it's uncertain if a different test design with 3D scrollable volumes could provide different results.

Furthermore, we evaluated the clinical value of memory-efficient CRF-GANs, but additional paradigms, such as variational auto-encoders [41] and normalizing flows [42], should be considered in future work.

# 5 Conclusion

Memory-efficient GANs, like CRF-GAN, offer significant advantages over traditional models, demonstrating comparable or even superior performance in both qualitative and quantitative evaluations. One of the key advantages of memory-efficient GANs is their ability to generate high-quality synthetic data despite having lower computational demands. In industries such as healthcare, where datasets are often limited, these GANs can step in to supplement existing data, effectively filling gaps that might otherwise go unaddressed. This becomes even more important when considering how expensive and time-consuming it is to manually collect and annotate rare medical cases. Moreover, by optimizing the way they utilize system resources, memory-efficient GANs can free up significant computational power. This saved power can be redirected towards other areas of model development, such as improving image quality or enhancing model performance.

# References

[1]  Miller D, Brown E: Artificial intelligence in medical practice: the question to the answer? *Am. J. Med*, 131(2):129–133, 2018.

[2]  Zheng Z, Wang M, Fan C, Wang C, He X, He X: Light&fast generative adversarial network for high-fidelity ct image synthesis of liver tumor. *Comput. Methods Programs Biomed,* 254:108-252, 2024.

[3]  Skandarani Y, Jodoin P-M, Lalande A: Gans for medical image synthesis: An empirical study. *J. Imaging*, 9(3):69, 2023.

[4]  Mendes J, Pereira T, Silva F, Frade J, Morgado J, Freitas C, Negrão E, De Lima B-F, Da Silva M-C, Madureira A, Ramos I, Costa J-L, Hespanhol V, Cunha A, Oliveira H-P: Lung ct image synthesis using gans. *Expert Syst Appl*, 215:119350, 2023.

[5]  Han C, Kitamura Y, Kudo A, Ichinose A, Rundo L, Furukawa Y, Umemoto K, Li Y, Nakayama H: Synthesizing diverse lung nodules wherever massively: 3d multi-conditional gan-based ct image augmentation for object detection. *International Conference on 3D Vision (3DV)*, 729–737. IEEE, 2019.

[6]  Goodfellow I, Pouget-Abadie J, Mirza M, Xu B, Warde-Farley D, Ozair S, Courville A, Bengio Y: Generative adversarial nets. *NIPS,* 2672-2680, 2014.

[7]  Shan H, Zhang Y, Yang O, Kruger U, Kalra M, Sun L, Cong W, Wang G: 3-d convolutional encoder-decoder network for low-dose ct via transfer learning from a 2-d trained network. *IEEE Trans Med Imaging*, 37(6):1522–1534, 2018.

[8]  Kudo A, Kitamura Y, Li Y, Iizuka S, Simo-Serra E: Virtual thin slice: 3d conditional gan-based super-resolution for ct slice interval. *Machine Learning for Medical Image Reconstruction (MLMIR), 91–100,* 2019.

[9]  Jin W, Fatehi M, Abhishek K, Mallya M, Toyota B, Hamarneh G: Artificial intelligence in glioma imaging: challenges and advances. *J. Neural Eng*, 17(2):021002, 2019.

[10]  Cirillo M, Abramian D, Eklund A: Vox2vox: 3d-gan for brain tumour segmentation. *Brainles2020,* 12658:274–284, 2021.

[11]  Dong X, Lei Y, Wang T, Thomas M, Tang L, Curran WJ, Liu T, Yang X: Automatic multiorgan segmentation in thorax CT images using U-net-GAN. *Med Phys*, 46(5):2157-2168, 2019.

[12]  Shubhangi N, Akshay D, Subrahmanyam M, Srivatsava N: RescueNet: An unpaired GAN for brain tumor segmentation. *Biomed Signal Process Control*, 55:101641, 2020.

[13]  Najafi H, Savoji K, Mirzaeibonehkhater M, Moravvej SV, Alizadehsani R, Pedrammehr S: A Novel Method for 3D Lung Tumor Reconstruction Using Generative Models. *Diagnostics*, 14(22):2604, 2024.

[14]  Chen C, Fu Z, Ye S, Zhao C, Golovko V, Ye S, Bai Z: Study on high-precision three-dimensional reconstruction of pulmonary lesions and surrounding blood vessels based on CT images. *Opt Express*, 32(2):1371-1390, 2024.

[15]  Swaminathan V-P, Balasubramani R, Parvathavarthini S, Gopal V, Raju K, Sivalingam, T-S, Rajan Thennarasu S: Gan based image segmentation and classification using vgg16 for prediction of lung cancer. *J. Adv. Res. Appl. Sci. Eng, 35*:45–61, 2024.

[16]  Frid-Adar M, Diamant I, Klang E, Amitai M, Goldberger J, Greenspan H: GAN-based synthetic medical image augmentation for increased CNN performance in liver lesion classification. *Neurocomputing,* 321:321-331, 2018.

[17]  Onishi Y, Teramoto A, Tsujimoto M, Tsukamoto T, Saito K, Toyama H, Imaizumi K, Fujita H: Automated Pulmonary Nodule Classification in Computed Tomography Images Using a Deep Convolutional Neural Network Trained by Generative Adversarial Networks. *Biomed Res Int*, 6051939, 2019.

[18]  Jin C-B, Kim H, Liu M, Jung W, Joo S, Park E, Ahn YS, Han IH, Lee JI, Cui X: Deep CT to MR synthesis using paired and unpaired data. *Sensors*, 19(10):2361, 2019.

[19]  Hong L, Modirrousta M-H, Nasirpour M-H, Chargari M-M, Mohammadi F, Moravvej S-V, Rezvanishad L, Rezvanishad M, Bakhshayeshi I, Alizadehsani R, Razzak I, Alinejad-Rokny H, Nahavandi S: GAN-LSTM-3D: An efficient method for lung tumour 3D reconstruction enhanced by attention-based LSTM. *CAAI Trans Intell Technol,* 1:1-24,2023.

[20]  Ferreira A, Pereira T, Silva F, Vilares A, Silva M, Cunha A, Oliveira H: Synthesizing 3d lung ct scans with generative adversarial networks. *Annu Int Conf IEEE Eng Med Biol Soc*: 2033–2036, 2022.

[21]  Sun L, Chen J, Xu Y, Gong M, Yu K, Batmanghelich K: Hierarchical amortized gan for 3d high resolution medical image synthesis. *IEEE J Biomed Health Inform*, 26(8):3966– 3975, 2022.

[22]  Lei Y, Wang T, Liu Y, Higgins K, Tian S, Liu T, Mao H, Shim H, Curran W, Shu H, Yang X: Mri-based synthetic ct generation using deep convolutional neural network. *Med Imaging 2019: Image Processing*,

10949:716–721, 2019.

[23] Yu B, Zhou L, Wang L, Fripp J, Bourgeat P: 3d cgan based cross-modality mr image synthesis for brain tumor segmentation. *IEEE 15th international symposium on biomedical imaging (ISBI 2018),* 626–630, 2018.

[24] Uzunova H, Ehrhardt J, Jacob F, Frydrychowicz A, Handels H: Multi-scale gans for memory-efficient generation of high resolution medical images. *Medical Image Computing and Computer Assisted Intervention–MICCAI 2019,* 11769:112-120, 2019.

[25] Gao J, Zhao W, Li P, Huang W, Chen Z: Legan: A light and effective generative adversarial network for medical image synthesis. *Comput Biol Med*, 148:105878, 2022.

[26] Holmes A-J, Hollinshead M-O, O'Keefe T-M, Petrov VI, Fariello G-R, Wald L-L, Fischl B, Rosen B-R, Mair R-W, Roffman J-L, Smoller J-W, Buckner R-L: Brain Genomics Superstruct Project initial data release with structural, functional, and behavioral measures. *Sci data*, 2:150031, 2015.

[27] Setio A-A-A, Traverso A, De Bel T, Berens M-S-N, Bogaard C-V-D, Cerello P, Chen H, Dou Q, Fantacci M-E, Geurts B, Gugten R-V, Heng P-A, Jansen B, de Kaste M-M-J, Kotov V, Lin J-Y, Manders JTMC, Sóñora-Mengana A, García-Naranjo J-C, Papavasileiou E, Prokop M, Saletta M, Schaefer-Prokop C-M, Scholten E-T, Scholten L, Snoeren M-M, Torres E-L, Vandemeulebroucke J, Walasek N, Zuidhof G-C-A, Ginneken B-V, Jacobs C: Validation, comparison, and combination of algorithms for automatic detection of pulmonary nodules in computed tomography images: the luna16 challenge. *Med image anal*, 42:1–13, 2017.

[28] Armato S-G 3rd, McLennan G, et al.: The lung image database consortium (LIDC) and image database resource initiative (IDRI): a completed reference database of lung nodules on CT scans. *Med Phys,* 38(2):915–931, 2011.

[29] Li Y, Ping W: Cancer metastasis detection with neural conditional random field. *ArXiv* 1806.07064, 2018.

[30] Gulrajani I, Ahmed F, Arjovsky M, Dumoulin V, Courville A-C: Improved training of Wasserstein GANs. *ArXiv* 1704.00028, 2017.

[31] Larsen A-B, Sønderby S-K, Larochelle H, Winther O: Autoencoding beyond pixels using a learned similarity metric. In: *International Conference on Machine Learning*, pages 1558–1566. PMLR, 2016.

[32] Kwon G, Han C, Kim D-S: Generation of 3D brain MRI using auto-encoding generative adversarial networks. In: *International Conference on Medical Image Computing and Computer-Assisted Intervention*, 118–126. Springer, 2019.

[33] Karras T, Aila T, Laine S, Lehtinen J: Progressive growing of GANs for improved quality, stability, and variation. *ArXiv* 1710.10196, 2018.

[34] Hong S, Marinescu R, Dalca A-V-D, Bonkhoff A-K, Bretzner M, Rost N-S, Golland P: 3D-StyleGAN: A style-based generative adversarial network for generative modeling of three-dimensional medical images. In: *Deep Generative Models, and Data Augmentation, Labelling, and Imperfections: First Workshop, DGM4MICCAI 2021, and First Workshop, DALI 2021, Held in Conjunction with MICCAI 2021, Strasbourg, France, October 1, 2021, Proceedings 1*, 24–34. Springer, 2021.

[35] Xing S, Sinha H, Hwang S-J: Cycle consistent embedding of 3D brains with auto-encoding generative adversarial networks. *Med Imaging with Deep Learning*, 2021.

[36] Pesaranghader A, Wang Y, Havaei M: CT-SGAN: Computed Tomography Synthesis GAN. *Deep Generative Models, and Data Augmentation, Labelling, and Imperfections. DGM4MICCAI DALI 2021. Lecture Notes in Computer Science,* vol 13003. Springer, Cham, 2021.

[37] Heusel M, Ramsauer H, Unterthiner T, Nessler B, Hochreiter S: GANs trained by a two time-scale update rule converge to a local Nash equilibrium. *Adv Neural Inf Process Syst*, 2017.

[38] Fortet R, Mourier E: Convergence de la répartition empirique vers la répartition théorique. *Annales Scientifiques de l'École Normale Supérieure* 70:267–285, 1953.

[39] Gretton A, Borgwardt K-M, Rasch M-J, Schölkopf B, Smola A: A kernel two-sample test. *J Mach Learn Res* 13(1):723, 2012.

[40] Huang G, Yuan Y, Xu Q, Guo C, Sun Y, Wu F, Weinberger K: An empirical study on evaluation metrics of generative adversarial networks. *ArXiv* 1806.07755, 2018.

[41] Jimenez Rezende D, Mohamed S, Wierstra D: Stochastic backpropagation and approximate inference in deep generative models. *International Conference on Machine Learning*, 1278–1286. PMLR, 2014.

[42] Dinh L, Sohl-Dickstein J, Bengio S: Density estimation using Real NVP. *ArXiv* 1605.08803, 2016.